\documentclass{PoS}

\PoS{PoS(LAT2005)090}

\usepackage{graphicx}
\usepackage{longtable}

\title{Electromagnetic mass difference on the lattice }

\ShortTitle{Electromagnetic mass difference on the lattice }

\author{\speaker{Yusuke Namekawa}\\
        Department of Physics, Nagoya University,\\
        Nagoya 464-8602, Japan\\
        E-mail: \email{namekawa@eken.phys.nagoya-u.ac.jp}}

\author{Yoshio Kikukawa\\
        Department of Physics, Nagoya University,\\
        Nagoya 464-8602, Japan\\
        E-mail: \email{kikukawa@eken.phys.nagoya-u.ac.jp}}

\abstract{
  We calculate
  electromagnetic mass difference of mesons using a method
  proposed by Duncan {\it et al}.
  The RG-improved gauge action and the non-compact Abelian gauge action
  are employed to generate configurations.
  Quark propagators in the range of $m_{PS}/m_{V}=0.76-0.51$
  are obtained with the meanfield-improved clover quark action.
  Chiral and continuum extrapolations are performed
  and the results are compared with experiments.
  Finite size effects are also examined.
  Quark masses are extracted from the measured spectrum.
  Our preliminary values for light quark masses are
  $m_{u}^{\overline{MS}}(\mu = 2~\mbox{GeV}) = 3.03(19)$~MeV,
  $m_{d}^{\overline{MS}}(\mu = 2~\mbox{GeV}) = 4.44(28)$~MeV,
  $m_{s}^{\overline{MS}}(\mu = 2~\mbox{GeV}) = 99.2(52)$~MeV.
}

\FullConference{XXIIIrd International Symposium on Lattice Field Theory\\
		25-30 July 2005\\
		Trinity College, Dublin, Ireland}

\begin{document}

\section{Introduction}
\label{section:Introduction}

Numerical simulations based on lattice QCD
allow us to calculate physical quantities
in high accuracy~\cite{plenary_lattice}.
However,
electromagnetic and isospin-violating effects
are usually ignored.
Taking into account these effects
is important for more realistic predictions.
In practice, it is pointed out that for light quark masses
systematic errors from electromagnetic effects
can be comparable with statistical one~\cite{HPQCD}.

An attempt was made to include electromagnetic effects
in lattice QCD~\cite{EM.Duncan}.
Pseudoscalar meson masses are calculated on the background of
gluon and photon fields.
$\pi^+ - \pi^0$ splitting is found to be $4.9(3)$~MeV
at $\beta=5.7$ with the Wilson action.
Their result shows a good agreement with experiments.
But their simulations were performed
at a single lattice spacing with unimproved actions.
The scaling violation of their result
is not expected to be small.
Another point is that the finite size corrections
were estimated only with a model.
It is desirable to evaluate finite size effects
from a first principle calculation directly.

In this work,
we study electromagnetic mass splittings of mesons
and extract quark masses
in the quenched theory.
\footnote{
For a similar work using $N_f=2$ DBW2 gauge and domain wall fermion
actions, we refer to Ref.~\cite{EM.Yamada}.
}
We employ an improved action combination,
the renormalization group(RG) improved gauge action
and the meanfield-improved clover quark action.
Pseudoscalar and vector meson masses
are computed at three lattice spacings
and extrapolated to the continuum limit.
Finite size effects are evaluated on lattices
with the spacial size of $L = 2.4$~fm and $L = 3.2$~fm.
Comparison of $m_{\pi}^+ - m_{\pi}^0$ as well as $m_{\rho}^+ - m_{\rho}^0$
with experimental values
are made.
Electromagnetic contribution to kaon mass splittings is also discussed.

\section{Method}
\label{section:method}

We generate $SU(3)$ and $U(1)$ fields
and calculate
quark propagators
on the combined $SU(3) \times U(1)$ configurations.
$SU(3)$ configurations are generated by the pseudoheat bath algorithm.
For the $U(1)$ gauge part
we employ a non-compact Abelian gauge action.
$U(1)$ configurations are constructed of the Fourier transformed
photon fields in the momentum space~\cite{EM.Dagotto}.
We fix the gauge to the Coulomb one in both $SU(3)$ and $U(1)$ parts
and use a gaussian smearing function.
Meson masses are computed from
quark propagators.
In this calculation, we use only connected part of quark propagators.
In the presence of electromagnetic fields,
isospin is no longer a conserved quantity.
Though the violation
for $m_{\pi^0}^2$ are expected to be small,
disconnected diagrams may give measurable contribution.
Estimate of the disconnected part is
under investigation.
Masses are obtained by $\chi^2$ fits to hadron correlators,
taking account of correlations among different time slices.
Statistical errors of hadron masses are estimated
with the jack-knife procedure.

Our simulations are performed
at $\beta=2.187$ ($a_{\rho} = 0.20$~fm), $\beta=2.334$ ($a_{\rho} = 0.16$~fm)
and $\beta=2.575$ ($a_{\rho} = 0.11$~fm)
using $12^{3} \times 24$, $16^{3} \times 32$ and $24^{3} \times 48$
lattices with the spatial extent $L \sim 2.5$~fm.
$16^{3} \times 24$ lattices at $\beta=2.187$ are also employed
to examine the finite size effects.
We take four hopping parameters
corresponding to $m_{PS}/m_V = 0.76-0.51$.
Configurations are generated
independently at each $m_{PS}/m_{V}$.
Measurements are carried out
at each $100$ sweeps.
Our simulation parameters are summarized
in Table~\ref{table:simulation_parameters}.
These parameters are chosen
so that they correspond to
those of two-flavor full QCD data generated by CP-PACS.

\begin{table}[tp]
\caption{Simulation parameters}
\label{table:simulation_parameters}
\leavevmode
\begin{center}
\begin{tabular}{l|ccc} \hline
$\beta$    & 2.187    & 2.334            & 2.575 \\
\hline
Size       & $12^3 \times 24$($16^3 \times 24$)
                      & $16^3 \times 32$ & $24^3 \times 48$ \\
\hline
$N_{conf}$ & 800(400) & 400              & 100 \\
\hline
\end{tabular}
\end{center}
\end{table}

\section{Simulation results}
\label{section:simulation_results}

\begin{figure}[tp]
  \includegraphics[width=75mm]{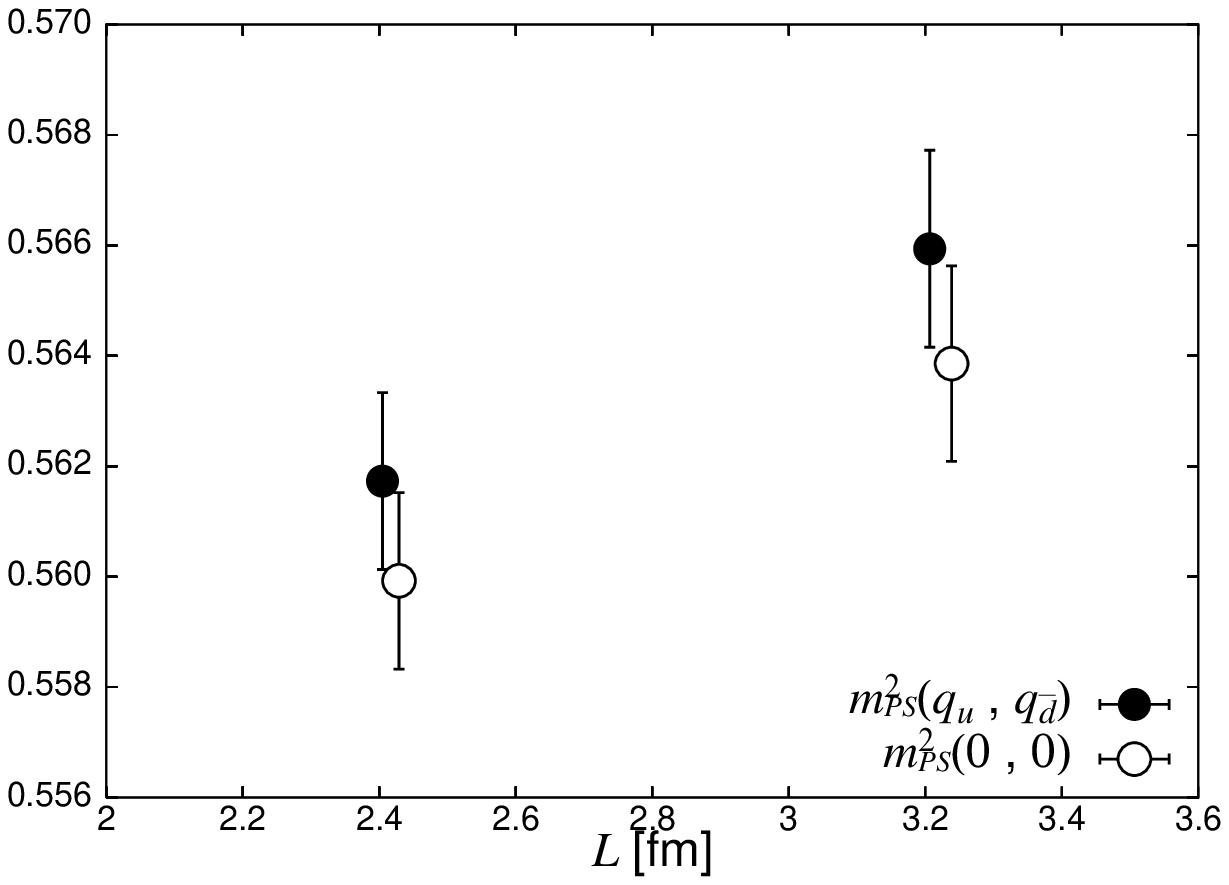}
  \includegraphics[width=75mm]{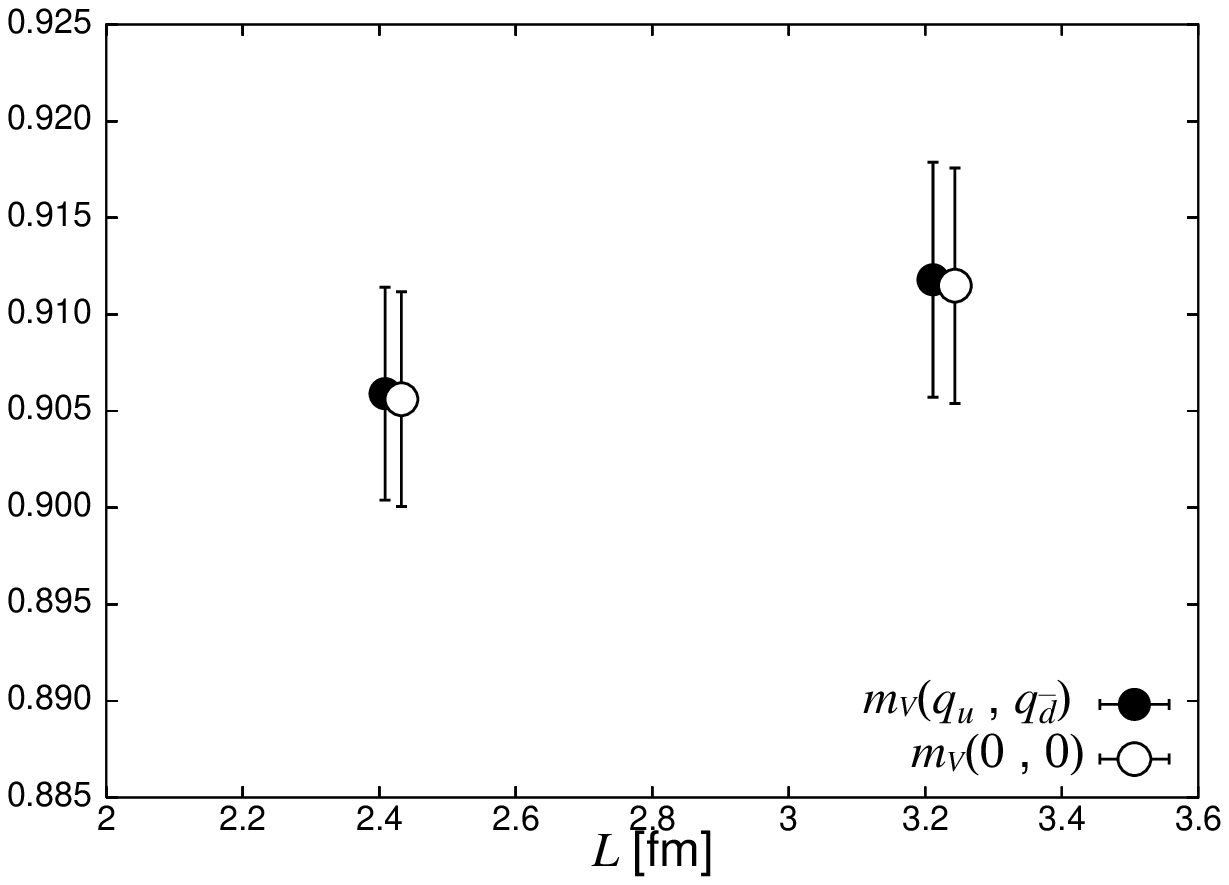}
\caption{
      Volume dependence of
      charged pseudoscalar (left panel) and vector meson masses (right panel)
      at $m_{PS}/m_{V} = 0.62$
      on $12^3 \times 24$ lattice.
      For comparison, masses in pure QCD (open symbols) are also plotted.
}
\label{figure:FSE_meson}
\end{figure}

We first check finite size effects in our results.
Finite size corrections may be enhanced
in the presence of electromagnetic fields
because electromagnetic fields have a long interaction range.
It is necessary to estimate magnitude of finite size corrections.
In Fig.~\ref{figure:FSE_meson},
we plot charged pseudoscalar and vector meson masses
as a function of the spatial volume.
The results obtained on $12^3 \times 24$ and $16^3 \times 24$ lattices
are mutually consistent within errors.
We did not find any enhancement of finite size effects
by electromagnetic fields.
In the quenched approximation,
$L=2.4$~fm seems to be enough for meson calculations,
even if there is an electromagnetic interaction.

\begin{figure}[tp]
  \includegraphics{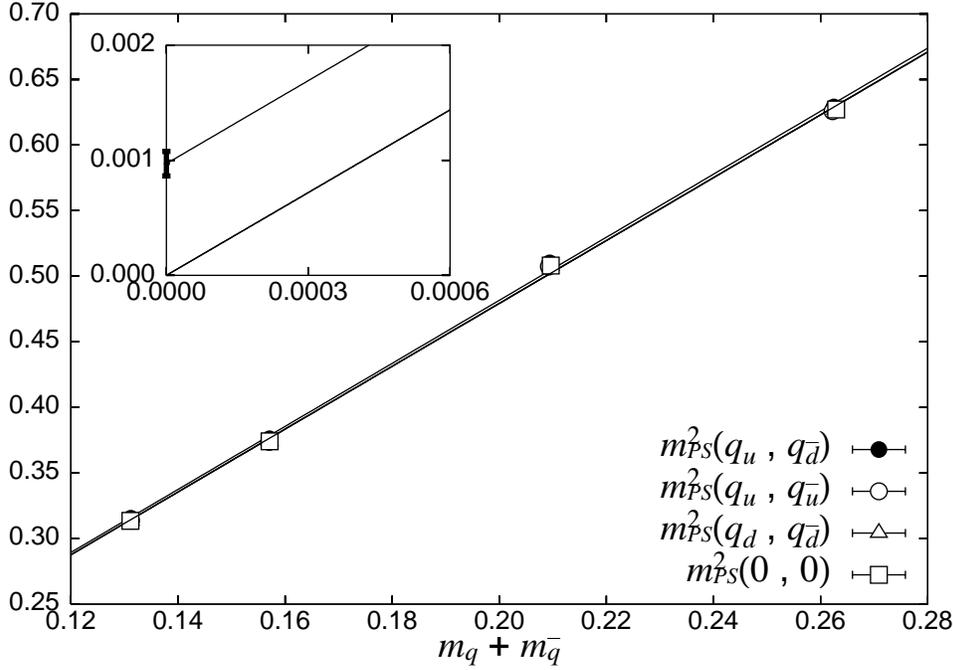}
\caption{
      Chiral extrapolations of
      pseudoscalar meson masses at $\beta = 2.187$.
}
\label{figure:m_u_m_d-m_PS2}
\end{figure}

In order to extrapolate our results to the chiral limit,
we fit a hadron mass as a function of quark masses and charges.
We employ the following form
for a chiral extrapolation of pseudoscalar meson masses.
\begin{eqnarray}
 m_{PS}^2 &=&  A_{PS}(q_{q} + q_{\overline{q}})^2
             + B_{PS}(q_{q},q_{\overline{q}}) (m_q + m_{\overline{q}}),
\label{equation:chiral_fit_to_m_PS2_type2}
\\
 &&B_{PS}(q_{q},q_{\overline{q}}) \equiv
   B^{PS}_0 + B^{PS}_2 (q_q + q_{\overline{q}})^2,
\end{eqnarray}
where $A_{PS}, B^{PS}_0, B^{PS}_{2}$
and $\kappa^c$ in $m_{q}, m_{\overline{q}}$
are fitting parameters.
$q_q$ is an electric charge for a particle
in units of $e$
and $q_{\overline{q}}$ is for an antiparticle.
$q = +2/3$ for the up quark and $q = -1/3$ for the down quark
are assigned.
We use a quark mass defined through a vector Ward identity,
\begin{eqnarray}
 m_{q} =
   \frac{1}{2}
   \left(   \frac{1}{\kappa_{q}}
          - \frac{1}{\kappa_q^{c}}
   \right), \ 
 m_{\overline{q}} =
   \frac{1}{2}
   \left(   \frac{1}{\kappa_{\overline{q}}}
          - \frac{1}{\kappa_{\overline{q}}^c}
   \right),
\end{eqnarray}
where $\kappa_{q}$ is a hopping parameter
for a quark and $\kappa_{\overline{q}}$ is for an antiquark.
$\kappa_{q,\overline{q}}^c$ are the corresponding critical
hopping parameters.
This extrapolation function,
inspired by the chiral perturbation theory,
only depends on the total charge and
the quark mass combination.
In the fits,
correlations among several charge combinations are neglected
for simplicity.
We estimate the errors by the jackknife method.
Strictly speaking,
the quenched chiral logarithm term must be added
to Eq.~(\ref{equation:chiral_fit_to_m_PS2_type2}).
However, the logarithmic curvature is not seen in our data
within $m_{PS}/m_V = 0.76 - 0.51$.
Smaller quark mass data are needed for a more precise extrapolation.
We found Eq.~(\ref{equation:chiral_fit_to_m_PS2_type2})
reproduce our lattice data well.
As an example,
the fit result at $\beta = 2.187$
is presented
in Fig.~\ref{figure:m_u_m_d-m_PS2}.
$\pi^+$ is heavier than $\pi^0$, as we see in nature.

Vector meson masses are extrapolated as follows.
\begin{eqnarray}
 m_{V} &=&  A_{V}(q_{q},q_{\overline{q}})
        + B_{V}(q_{\overline{q}},q_{d}) (m_q + m_{\overline{q}}),
\label{equation:chiral_fit_to_m_V}
\\
 &&A_{V}(q_{q},q_{\overline{q}}) \equiv
   A^{V}_0 + A^{V}_2 (q_q + q_{\overline{q}})^2,\\
 &&B_{V}(q_{q},q_{\overline{q}}) \equiv
   B^{V}_0 + B^{V}_2 (q_q + q_{\overline{q}})^2.
\end{eqnarray}
$A^{V}_0, A^{V}_2, B^{V}_0$
and $B^{V}_2$ are fitting parameters.
In contrast to the case of pseudoscalar mesons,
mass difference of charged and neutral vector mesons
is found to be small.
This tiny mass difference
is consistent with the result calculated
by the hidden local symmetry formulation,
$\Delta m_{\rho} = -1$~MeV~\cite{delta_m_rho.Harada}.

We identify the physical point with
experimental values of $\pi^0$ and $\rho^0$ masses,
$M_{\pi^0}=0.1350$~GeV and $M_{\rho^0}=0.7751$~GeV.
By solving Eq.~(\ref{equation:chiral_fit_to_m_PS2_type2})
and Eq.~(\ref{equation:chiral_fit_to_m_V})
using $M_{\pi^0}, M_{\rho^0}$,
a sum of bare up and down quark masses
at the physical point, $(m_u + m_d)_{phys}$, are determined.
The lattice spacing $a_{\rho}$ is set by identifying $M_{\rho^0}$
with $m_{V}$.
Substituting $(m_u + m_d)_{phys}$
to Eq.~(\ref{equation:chiral_fit_to_m_PS2_type2})
predicts $\pi^+$ mass.
Similarly, Eq.~(\ref{equation:chiral_fit_to_m_V}) gives
$m_{\rho^+}$.
To determine up, down and strange quark masses independently,
we also use experimental values of $K_+$ and $K_0$ meson masses as inputs.
$(m_u + m_s)_{phys}$ is obtained from $M_{K_+} = 0.4937$~GeV
and $(m_d + m_s)_{phys}$ from $M_{K_0} = 0.4976$~GeV.
Quark masses are renormalized
using one-loop renormalization constants $Z_m$
and coefficients $b_m$ at $\mu = 1/a$~\cite{Z-factor},
\begin{equation}
 m_{q}^{\overline{\rm MS}}(\mu = 1/a)
 = Z_m \left( 1 + b_m \frac{m_{q}}{u_0} \right)
                      \frac{m_{q}}{u_0},
 \label{equaton:renormalized_m_quark_VWI}
\end{equation}
where $u_0$ is a tadpole factor, $u_0 = (1 - 0.8412 / \beta)^{1/4}$.
The $\overline{\mbox{MS}}$ quark masses at $\mu = 1/a$
are evolved to $\mu = 2$~GeV
using the four-loop
beta function.

\begin{figure}[tp]
  \includegraphics[width=75mm]{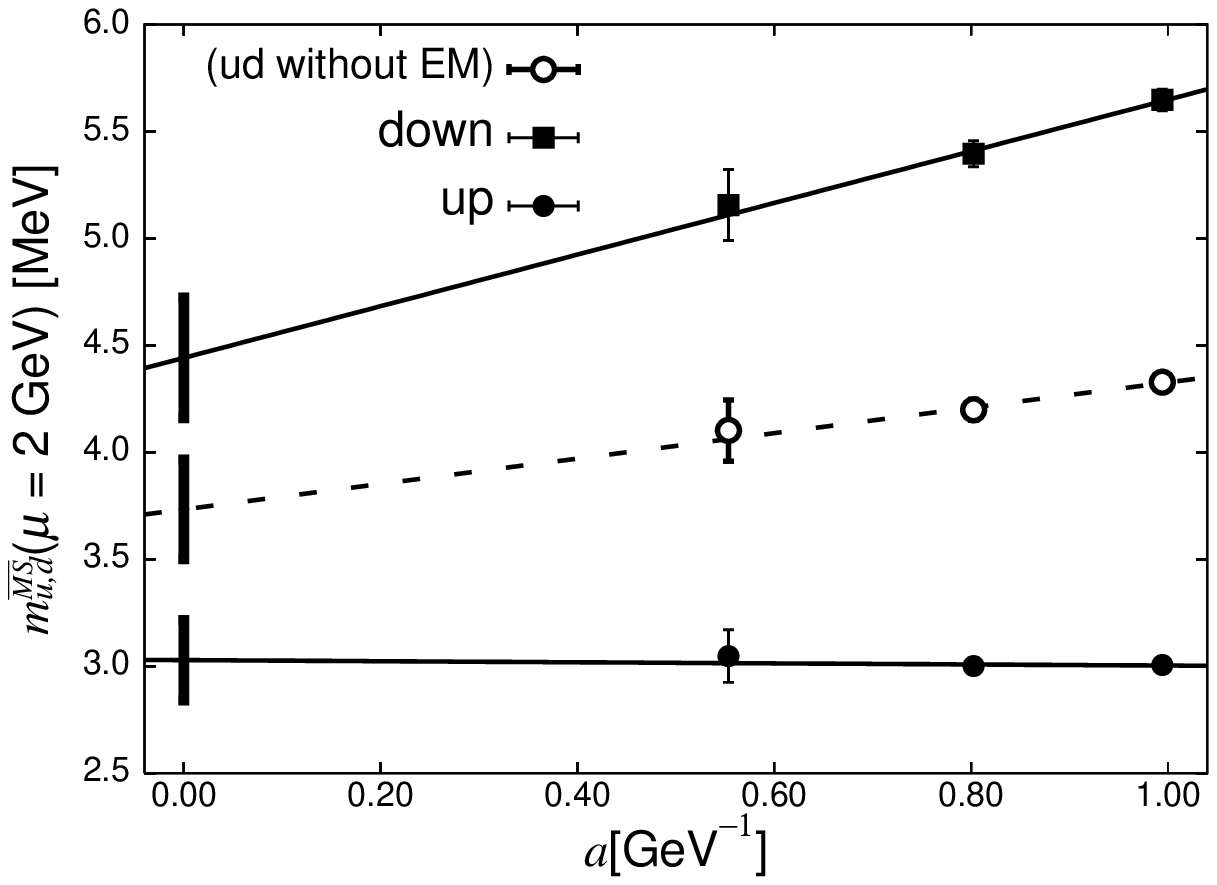}
  \includegraphics[width=75mm]{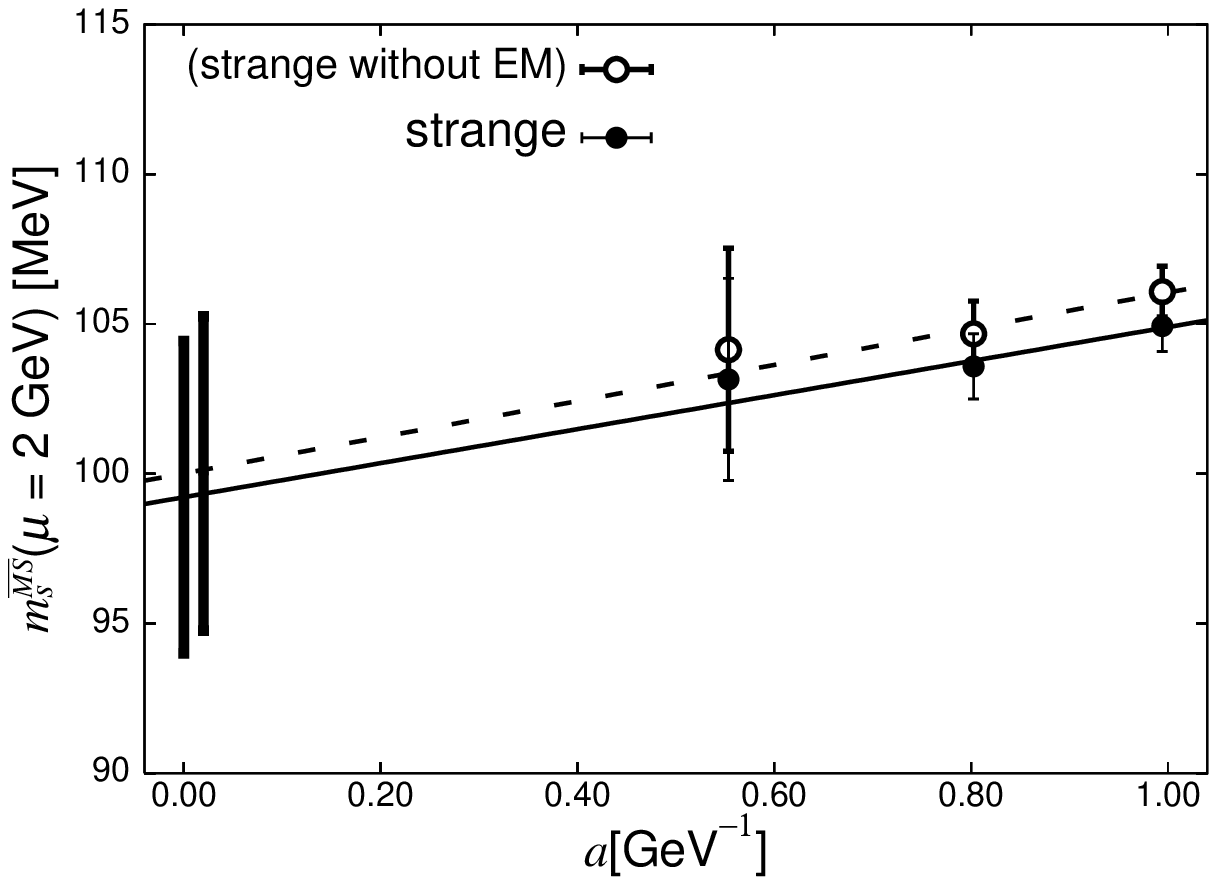}
\caption{
      Lattice spacing dependence of
      quark masses with and without electromagnetic effects.
}
\label{figure:a_inv-m_uds}
\end{figure}

\begin{figure}[tp]
  \includegraphics[width=75mm]{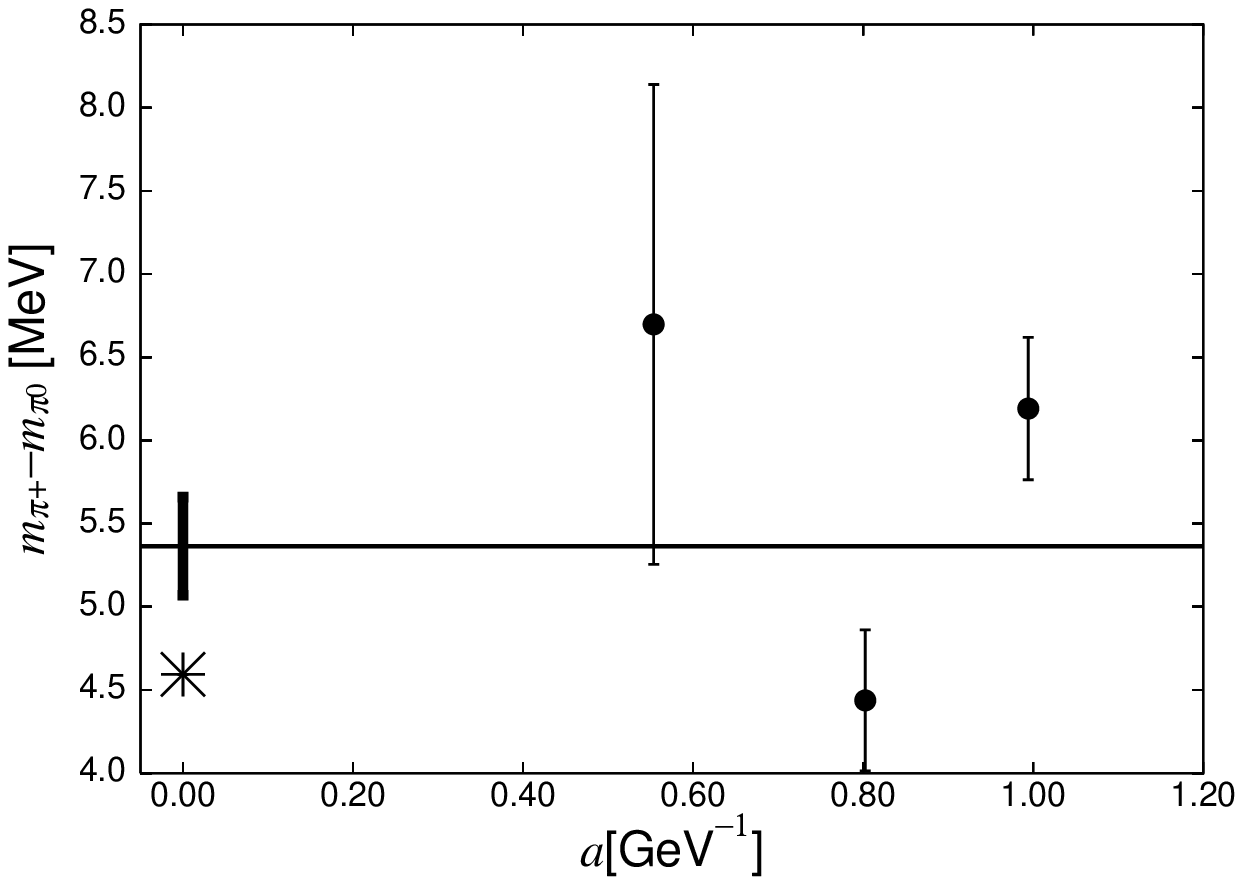}
  \includegraphics[width=75mm]{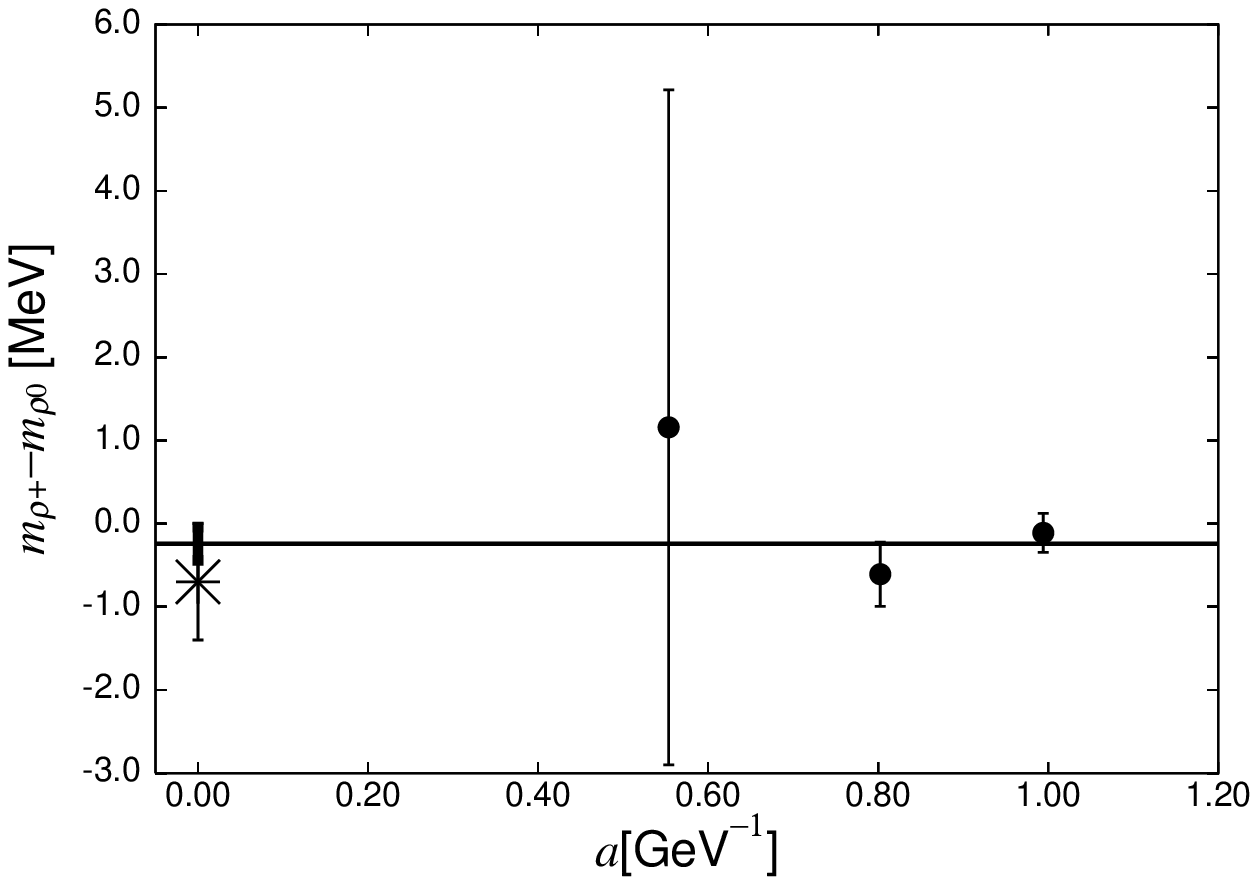}
\caption{
      Lattice spacing dependence of
      electromagnetic mass splittings.
      Star represents the experimental value.
}
\label{figure:a_inv-delta_m}
\end{figure}

Continuum extrapolations are performed
by linear fits to the data at three lattice spacings.
Our preliminary results for light quark masses are
$m_{u}^{\overline{MS}}(\mu = 2~\mbox{GeV}) = 3.03(19)$~MeV,
$m_{d}^{\overline{MS}}(\mu = 2~\mbox{GeV}) = 4.44(28)$~MeV,
$m_{s}^{\overline{MS}}(\mu = 2~\mbox{GeV}) = 99.2(52)$~MeV.
Fig.~\ref{figure:a_inv-m_uds} illustrates
lattice spacing dependence of quark masses
with and without electromagnetic effects.
We found electromagnetic contributions to the strange quark mass
is 1\%.
In contrast to the case of quark masses,
electromagnetic mass splittings of pseudoscalar and vector mesons
show mild lattice spacing dependence.
Therefore, we employ constant fits
for continuum extrapolations of
electromagnetic mass splittings.
Our results are represented in Fig.~\ref{figure:a_inv-delta_m}.
The obtained mass splittings are
consistent with experimental values.
In addition to $\pi^+ - \pi^0$ and $\rho^+ - \rho^0$
mass difference,
we can estimate electromagnetic contribution
to $K^+ - K^0$ mass difference by
Eq.~(\ref{equation:chiral_fit_to_m_PS2_type2}).
Using a constant continuum extrapolation again,
the electromagnetic contribution to $K^+ - K^0$ mass difference
is evaluated to be $1.420(24)$~MeV,
which is close to the value of
Dashen's theorem $1.3$~MeV~\cite{EM.Dashen},
rather than a model estimate $2.6$~MeV~\cite{EM.Bijnens}.
But, our simulations are in the quenched approximation
and the chiral logarithm is neglected.
Including dynamical quarks and chiral logarithm effects
is needed for a more precise comparison with other calculations.

\section{Conclusions}
\label{section:conclusion}

We calculated electromagnetic mass splittings
using the RG-improved gauge action and the clover-improved
Wilson quark action
on the background of $SU(3) \times U(1)$ fields
in the quenched approximation.
After chiral and continuum extrapolations,
we found $\pi^+ - \pi^0$ and $\rho^+ - \rho^0$ mass differences
are consistent with experimental values,
and we can extract up, down and strange quark masses.
We also confirmed that mass differences obtained with $L = 2.4$~fm
are not shifted by finite size effects.
An important future work is to include
dynamical quark effects.
Simulations using full QCD data
generated by CP-PACS collaboration
are ongoing.

\begin{acknowledgments}
We thanks M.~Harada, M.~Hayakawa, M.~Tanabashi, N.~Yamada and K.~Yamawaki
for valuable discussions.
A part of our calculations have been carried out
on a Hitachi SR8000 at KEK,
a cluster machine at Center for Computational Sciences,
University of Tsukuba,
and a supercomputer (NEC SX-5) at Research Center
for Nuclear Physics,
Osaka University.
This work is supported in part
by $21^{\mbox{st}}$ Century COE Program, Nagoya University.
\end{acknowledgments}


\begin{thebibliography}{99}

\bibitem{plenary_lattice}
For a recent review, see
T.~Izubuchi,
these proceedings.


\bibitem{HPQCD}
C.~Aubin {\it et al.},
Phys. Rev. D {\bf 70} 031504 (2004);
Phys. Rev. D {\bf 70} 114501 (2004).


\bibitem{EM.Duncan}
A.~Duncan {\it et al.},
Phys. Rev. Lett. {\bf 76} 3894 (1996).


\bibitem{EM.Yamada}
N.~Yamada {\it et al.},
these proceedings.


\bibitem{EM.Dagotto}
E.~Dagotto {\it et al.},
Nucl .Phys. B {\bf 331} 500 (1990).


\bibitem{delta_m_rho.Harada}
M.~Harada and K.~Yamawaki,
Phys. Rept. {\bf 381} 1 (2003).


\bibitem{Z-factor}
S.~Aoki, K.~Nagai, Y.~Taniguchi and A.~Ukawa,
Phys. Rev. D {\bf 58}, 074505 (1998);\\
Y.~Taniguchi and A.~Ukawa,
{\it ibid.} {\bf 58}, 114503 (1998).\\
S.~Aoki, R.~Frezzotti and P.~Weisz,
Nucl. Phys. B {\bf 540}, 501 (1999).


\bibitem{EM.Dashen}
R.~Dashen,
Phys. Rev. {\bf 183} 1245 (1969).


\bibitem{EM.Bijnens}
J.~Bijnens,
Phys. Lett. B {\bf 306} 343 (1993).


\end{thebibliography}
\end{document}